\documentstyle[12pt]{article} 
\newcommand{\be}{\begin{equation}}
\newcommand{\ee}{\end{equation}}
\newcommand{\ba}{\begin{array}}
\newcommand{\ea}{\end{array}}
\newcommand{\bc}{\begin{center}}
\newcommand{\ec}{\end{center}}

\newcommand{\plb}[1]{Phys. Lett. {\bf #1}}

\newcommand{\prl}[1]{ Phys. Rev. Lett. {\bf #1}}

\newcommand{\apb}[1]{Ann. of Phys. (N.Y.) {\bf #1} }
\newcommand{\npb}[1]{Nucl. Phys. {\bf #1}}

\begin{document}

\title{
Ground State Properties and Glueball Spectrum in SU(2) Yang-Mills Theory \\
using Gauge Invariant Variables
}
\author{
C\'ecile Martin
and
Dominique Vautherin \\
\\
{\normalsize
Division de Physique Th\'eorique\thanks{Unit\'e de Recherche des Universit\'es
Paris XI et Paris VI associ\'ee au C.N.R.S},}\\
{\normalsize  Institut de Physique Nucl\'eaire,} \\
{\normalsize F-91406 , Orsay Cedex, France }
}

\date{}

\maketitle
\vspace*{1cm}
\begin{abstract}
 We describe a nonperturbative calculation of the spectrum of SU(2)
Yang-Mills theory based on a Hamiltonian formulation. Our approach
exploits gauge invariant variables similar to those used in nuclear physics
to describe collective motion in nuclei.
\end{abstract}


\noindent IPNO/TH 93-68
\newpage

 Our understanding of the low energy behaviour of QCD depends crucially on
the developpment of nonperturbative methods. The variational gaussian
approximation which has been successfull in quantum mechanics and in
scalar field theory remains difficult to apply for a nonabelian gauge
theory because of the requirement to satisfy the Gauss law constraint.
The necessity to maintain gauge invariance is however an essential
ingredient in practical calculations.
In order to study the configurations which contribute significantly to
the nonperturbative ground state, we have to take into account  the
gauge invariant functional measure. The nontrivial gauge invariant volume
element will induce a centrifugal effect and the boundary conditions on the
wave functional will differ drastically from those
in  perturbative calculations.
This has important phenomelogical
consequences such as the occurence of nonvanishing condensates
describing the vacuum state and the existence of a mass gap \cite{1}.

 In this paper, we use a representation of the SU(2) vector potential
which separates explicitly the gauge degrees of freedom \cite{2,3}.
In this representation, the Gauss law appears as a  local constraint.
The hamiltonian
however becomes non-local. A derivative expansion
(or strong coupling expansion) valid in the nonperturbative domain and for
slowly-varying fields allows one to write expliciltly the first few  terms of
an effective hamiltonian. One can thus obtain approximate solutions for the
dynamics which are gauge invariant.

In section 2, we introduce more appropriate gauge invariant variables
$\rho(x), \beta(x), \gamma(x)$. These variables describe the field
configuration in an intrinsic frame and they can be interpreted as
"density" and "deformation" variables. They are analogous to the
collective variables introduced by Bohr and Mottelson to describe the
dynamics of deformed nuclei \cite{4,5}, the pure gauge degree of freedom
corresponding to intrinsic nucleon coordinates.
The collective gauge invariant variables $\rho, \beta, \gamma$
allow one to describe deformed solutions
of the dynamical equation without breaking the local gauge symmetry
of the hamiltonian.

In section 3 and 4,
we work in  the strong coupling limit
which corresponds to the constant field approximation. We give the expression
of the hamiltonian in terms of the gauge invariant variables $\rho, \beta,
\gamma$.
We investigate the properties of the ground
state in the
SU(2) gauge theory and we obtain predictions for the lowest masses of
the color singlet bound states of gluons called glueballs.
The minimum of the energy corresponds to a configuration with axial symmetry
($\gamma=0$) and a strong deformation ($\beta$ near $1$).
In this formalism, the lowest glueball states are interpreted as rotational
levels and vibrational collective levels.
In the last section, we discuss our results for the behaviour of the
wavefunctional.

\section{The polar representation}

  In the hamiltonian formalism, we choose the temporal gauge $A_0^{a}=0$.
The hamiltonian reads :
\be \label{1e1}
H=\frac{1}{2} \int \: d^3x \: tr \left( E^2 + B^2 \right) \ . \ee
For the SU(2) color group, the vector potential $A_{ia}$ (where
i is a space index
and a is a color index ) is a $3 \times 3$
matrix. The polar representation is given by \cite{2,3} :
\be \label{1e2}
A_{ia}=f_{in} \lambda_{n} h_{na} - \frac{1}{2g} h_{kb} \partial_{i} h_{kc}
\epsilon_{abc} \ , \ee
where $\lambda_{n}, n=1,2,3$ are three numbers, $\hat f(\theta_{i})$ and
$\hat h(\phi_a)$ are orthogonal matrices parametrized by two sets of three
Euler angles $\theta_{i}$ and $\phi_a$ and $g$ is the bare coupling constant.
In terms of the $3 \times 3$ spin 1 matrices $(S_i)_{jk}=-i \epsilon_{ijk}$,
\be \label{1e2a}
\hat f =\exp \left(-i \theta_1  S_z\right)
\: \exp \left( -i \theta_2 S_y \right)
\: \exp \left( -i \theta_3 S_z \right) \ . \ee
The matrix $\hat f$ describes a rotation
in ordinary space and the matrix $\hat h$ a rotation in color space.
 In general, $\lambda_n, \theta_i$ and
$\phi_{a}$ are space-dependent. Under a local gauge transformation,
the matrix $\hat h$
is simply rotated while $\hat f$ and $\lambda_{n}$ remain unchanged.
Therefore, among the nine
variables $A_{ia}$, six gauge invariant variables $\lambda_{n}$ and
$\theta_i$ are explicitly separated from the three gauge degrees of freedom
$\phi_a$.

In these new variables, the Gauss law operator
\be \label{1e3}
G^l(x) \equiv \frac{1}{g} \: D_i^{la} E_{ia} \ , \ee
becomes a local operator :
\be \label{1e4}
G^l(x)=J^l(x) \ , \ee
where $J^l$ are the cartesian components of the color angular momentum in the
laboratory frame expressed in terms of the Euler angles $\phi_a$  :
\be \label{1e4a}
J_1= -i \left\{ -\cos \phi_1 \cot \phi_2 \frac{\partial}{\partial \phi_1}
- \sin \phi_1 \frac{\partial}{\partial \phi_2}+ \frac{\cos \phi_1}{\sin \phi_2}
\frac{\partial}{\partial \phi_3} \right\}   \ , \ee
with similar formulae for $J_2$ and $J_3$ \cite{8a}.
For a color singlet state $\vert \Psi >$
\be \label{1e5}
G^l(x) \vert \Psi > = J^l(x) \vert \Psi > = 0 \ . \ee
Therefore, the wave functional for a color singlet state depends only on
the six gauge invariant variables : $\Psi (\lambda_{n}(x), \theta_i(x))$.
The functional integration measure becomes
\be \label{1e6}
\prod_{i,a} D A_i^a (x) = \prod_{n>m} \left \vert \lambda_n^2(x) -
\lambda_m^2(x) \right \vert \: \prod_{p} D \lambda_p(x) \:
d \hat f \:  d \hat h \ ,  \ee
where $d \hat f = \sin \theta_2 \: d \theta_1 d \theta_2 d \theta_3$, and
$d \hat h = \sin \phi_2 \: d \phi_1 d \phi_2 d \phi_3$.

 In terms of the new gauge invariant variables, the hamiltonian becomes
non local. The expression for the potential energy $\frac{1}{2}B^2$
is given by :
\be \label{1e7}
V = \frac{g^2}{2} (\lambda_1^2 \lambda_2^2 + \lambda_2^2 \lambda^2_3 +
\lambda^2_1 \lambda^2_3) + \frac{1}{8} \sum_n \left(
curl \: \vec w_n \right )^2
\ ,  \ee
where $\vec w_n$ are the three orthogonal vectors : $w^i_n= f_{in} \lambda_n$.

The kinetic energy is a non local function of the gauge invariant quantities
and their derivatives. However, we can use a derivative (or strong coupling)
expansion to write explicitly the first few terms.
This expansion will be valid in the nonperturbative domain and
for slowly-varying fields.
In the following, we will consider only the lowest order in $1/g^2$,
which corresponds also to the constant field approximation.
At this order,
we keep only the constant term in (\ref{1e2}) , which is the non-abelian term.
This approach is therefore not adequate to describe the abelian limit.

For color singlet states,
the kinetic energy is given by \cite{2}
\be \label{1e8} \ba{ll}
T = & {\displaystyle
 - \frac{1}{2 L^3} \left\{  \sum_{n=1}^{3}
 \frac{\delta^2}{\delta \lambda_n \delta \lambda_n }    +
2  \sum_{n>m} \frac{1}{\lambda_n^2 - \lambda_m^2} \left(
\lambda_n \frac{\delta}{\delta \lambda_n} - \lambda_m \frac{\delta}
{\delta \lambda_m} \right) \right\} } \\
& {\displaystyle
   + \frac{1}{4 L^3}
\sum_{k,n,m} \frac{\lambda_n^2 + \lambda_m^2}{(\lambda_n^2 -
\lambda_m^2)^2} \epsilon^2_{nmk} L_k^2
\ , } \ea  \ee
We have introduced a length scale L, the total volume being $L^3$.
The operators
 $L_k$ are the components of the angular momentum in the proper frame.
As an example \cite{8a} :
\be \label{1e8a}
L_x= -i \left\{ - \frac{\cos \theta_3}{\sin \theta_2} \: \frac{\partial}
{\partial \theta_1} + \sin \theta_3 \frac{\partial}{\partial \theta_2} +
\cot \theta_2 \cos \theta_3 \frac{\partial}{\partial \theta_3} \right\}
\ . \ee
In  lowest order in $1/g^2$, the angular momentum in the laboratory
$M_i= f_{ik} L_k$ is equal to the spin density :
$\vec M (x) = \vec A^a(x) \times \vec E^a (x)$.

It can be also usefull to write the expression of the gauge invariant
operator $\vec B^a . \vec E^a$. For color singlet states, it is given by :
\be \label{1e9}
\vec B^a . \vec E^a = - i \: g \: \epsilon_{npq} \: \epsilon_{mpq} \:
\lambda_p
\lambda_q \left [ \delta_{nm} \frac{\delta}{\delta \lambda_m} +
i \: \frac{\epsilon_{knm}}{\lambda_n^2-\lambda^2_m} \lambda_m L_k \right ]
\ . \ee

\section{The gauge invariant collective coordinates $\rho(x), \beta(x),
\gamma(x)$ }

 States of zero angular momentum depend only on three gauge invariant and
rotational invariant quantities which can be choosen as \cite{9} :
\be \label{2e1}
B^2=B_i^a(x) B_i^a(x) \ ,  \ee
\be \label{2e2}
B.y=B_i^a(x) y_i^a(x) \ , \ee
and
\be \label{2e3}
y^2=y_i^a(x) y_i^a(x) \ , \ee
where $y_i^a (x) =\epsilon^{abc} \epsilon_{ikl} B_k^b(x) B_l^c(x)$.
For constant fields, they are related to the gauge invariant variables
$\lambda_n$ according to the formulae :
\be \label{2e4}
B^2=g^2 \left( \lambda_1^2 \lambda_2^2 + \lambda_2^2 \lambda_3^2 +
\lambda_1^2 \lambda_3^2 \right) \ , \ee
\be \label{2e5}
y^2 = 4 g^4 \lambda_1^2 \lambda_2^2 \lambda_3^2 \: \left(
\lambda_1^2 + \lambda_2^2 + \lambda_3^2 \right) \ , \ee
\be \label{2e6}
B . y = 6 g^3 \lambda_1^2 \lambda_2^2 \lambda_3^2 \ . \ee
These expresions, as well as the functional integration measure (\ref{1e6}),
are invariant under a permutation of the $\lambda_n$ and a simultaneous
 change of sign
of two $\lambda_n$.

In the following, we will use three gauge invariant variables $\rho, \beta,
\gamma$ defined as :
\be \label{2e7}
\rho^2 = \lambda_1^2 + \lambda_2^2 + \lambda_3^2 \ , \ee
\be \label{2e8}
\lambda_2^2 - \lambda_1^2 = \frac{2}{\sqrt 3} \rho^2 \beta \sin \gamma \ , \ee
\be \label{2e9}
2 \lambda_3^2 - \left( \lambda_1^2 + \lambda_2^2 \right) = 2 \rho^2 \beta
\cos \gamma \ , \ee
i.e.,
\be \label{2e9a}
\lambda_1^2 - \frac{\rho^2}{3} = \frac{2}{3} \rho^2 \beta \cos
\left( \gamma + \frac{2 \pi}{3} \right) \ . \ee
\be \label{2e9b}
\lambda_2^2 - \frac{\rho^2}{3} = \frac{2}{3} \rho^2 \beta \cos
\left( \gamma - \frac{2 \pi}{3} \right) \ . \ee
\be \label{2e9c}
\lambda_3^2 - \frac{\rho^2}{3} = \frac{2}{3} \rho^2
\beta \cos  \gamma   \ . \ee
where $0<\beta<1$ and, from symmetry properties,  we can restrict the angle
$\gamma$ between 0 and $\pi/3$.  The variable $\rho$ has the dimension of
$1/L$.

The vibrational part of the wave function ( i. e. the part of the wave function
which is independent of the three Euler angles $\theta_i$) is a function of
the three gauge invariant collective coordinates $\rho, \beta, \gamma$ :
$\Psi \left( \rho, \beta, \gamma \right)$.
The variables $\rho, \beta, \gamma$ are interpreted as the analogous of
the density and deformation variables used in the collective model of
Bohr  in
Nuclear Physics \cite{4}.
They remind also the derivation of the collective hamiltonian for a
system of N particules by introducing 3 N $-$ 9 Euler angles and
six collective coordinates which describe the shape, the dimension
and the orientation of the system \cite{5}. Since we start from nine
degrees of freedom $A_i^a$ for the Yang-Mills system, the description
in terms of the gauge invariant variables corresponds to the treatment
of the four-body problem \cite{5}. The variables $\rho, \beta, \gamma$
will describe  the gluon configuration
in the intrinsic frame.
The $\rho$ vibrations correspond to density vibrations of monopole
character. The $\beta$ and $\gamma$ vibrations correspond to
quadrupole oscillations. In general, there will be a coupling between
the oscillations of the density $\rho$ and those of the deformations
$\beta$, $\gamma$.
We will show that $\rho, \beta$ and $\gamma$ are convenient coordinates
to perform practical calculations. Furthermore, they give a physical insight in
the structure of the vacuum state and the lowest excited states.

In the  $\beta, \gamma$ plane, the axis $\beta =0$ corresponds to a
"spherical" field configuration : $\lambda_1 = \lambda_2 = \lambda_3$.
The axis $\gamma = 0$ corresponds to an "axial symmetric" field
configuration :
\be
\lambda_1^2 = \lambda_2^2 = \frac{\rho^2}{3} \: \left( 1 - \beta^2 \right)
\ , \ee
\be
\lambda_3^2 = \frac{\rho^2}{3} \: \left( 1 + 2 \beta \right)
 \ . \ee
The point $\beta = 1, \gamma = 0$ corresponds to two vanishing $\lambda$'s :
$\lambda_1 = \lambda_2 = 0$, and $ \lambda_3^2 = \rho^2$. We point out that
 this
configuration does not  describe an abelian type field. Indeed, in the strong
coupling approximation which corresponds to the constant potential
approximation, we are not in a position to investigate the abelian limit.

An arbitrary point in the $\beta, \gamma$ plane corresponds to a "triaxial"
field configuration.

The gauge invariant quantities in eq. (\ref{2e4}-\ref{2e6}) become  functions
of $\rho^2, \beta^2$ and $\beta^3 \cos 3\gamma$ :
\be \label{2e10}
\frac{B^2}{2} = g^2 \frac{\rho^4}{6} (1- \beta^2) \ , \ee
\be \label{2e11}
B . y = g^3 \: \frac{2}{3} \: \rho^6 \left( \frac{1}{3} - \beta^2 + \frac{2}{3}
\beta^3 \cos 3 \gamma \right) \ ,  \ee
and the expression for the volume element of the vibrational coordinates
\be \label{2e12}
d \tau = \vert \lambda_1^2 -\lambda_2^2 \vert \vert \lambda_2^2 - \lambda_3^2
\vert \vert \lambda_1^2 - \lambda_3^2 \vert  \: d \lambda_1 \: d\lambda_2
\: d \lambda_3 \ , \ee
becomes :
\be \label{2e13}
d \tau = \frac{2}{3 \sqrt 3} \: \rho^6 \beta^3 \vert \sin 3 \gamma \vert
\: \vert \det M \vert \: d \rho \: d \beta \: d \gamma \ , \ee
where the jacobian is given by
\be
\det M \equiv
\frac{D \left(  \lambda_1 \lambda_2 \lambda_3 \right)}{D \left(
\rho \: \beta \: \gamma \right)} = \frac{1}{6 \sqrt 3 \lambda_1 \lambda_2
\lambda_3} \: \rho^5 \beta \ . \ee

A singular point occurs when one of the $\lambda_n$ vanishes or when two
$\lambda$'s are equal. This will imply
suitable boundary conditions on the gauge invariant wave function
$\Psi \left( \rho, \beta, \gamma \right)$. Let us now
 use the following rescaling :
\be \label{2e14}
\Phi \left(\rho, \beta, \gamma \right) = \rho^{11/2} \:
\frac{1}{\left\vert \lambda_1 \lambda_2 \lambda_3 \right\vert^{1/2}} \:
\Psi \left( \rho, \beta, \gamma \right) \ . \ee
The wave function $\Phi$ is normalized according to the integration measure
\be \label{2e15}
\beta^4 \: \vert \sin 3 \gamma \vert \: d \rho \: d \beta \: d \gamma \ . \ee

\section{Strong coupling expansion}

In terms of the  new gauge invariant variables $\rho, \beta, \gamma$ and
after the rescaling (\ref{2e14}), the vibrational hamiltonian (i.e.
the terms in  equations  (\ref{1e7}) and (\ref{1e8}) independent of the
angular momentum operators $L_k$ )
is equal to  $ T_{vib} + V'$ with
\be \label{3e1}
\ba{llll}
T_{vib} = - \frac{1}{2 L^3}
& {\displaystyle \left\{
\frac{\partial^2}{\partial \rho^2} \right. } \\
& {\displaystyle \left.
+ \frac{2}{\rho^2} \left( 1 + \beta \cos 3 \gamma - 2 \beta^2 \right)
\frac{\partial^2}
{\partial \beta^2} + \frac{2}{\rho^2 \beta^2} \left( 1 - \beta
\cos 3 \gamma \right) \frac{\partial^2}{\partial \gamma^2} \right. } \\
& {\displaystyle \left.
- \frac{4}{\rho^2} \sin 3 \gamma \frac{\partial^2}
{\partial \beta \partial \gamma}
+ \frac{2}{\rho^2 \beta} \left( 4 - \beta \cos 3 \gamma -  12 \beta^2 \right)
\frac{\partial}{\partial \beta} \right. } \\
& {\displaystyle \left.
+ \frac{2}{\rho^2 \beta^2 \sin 3 \gamma} \left( 3 \cos 3 \gamma - 3 \beta
+ 2 \beta \sin^2 3 \gamma \right) \frac{\partial}{\partial \gamma} \right\}
\ , }
\ea \ee
and
\be \label{3e2}
V'\left( \rho, \beta, \gamma \right)
=L^3 \: \frac{g^2}{6} \rho^4 \left( 1-\beta^2 \right) +
\frac{99}{8 L^3 \rho^2} + \frac{3}{8 L^3 \rho^2} \:
\frac{1- \beta^2}{\frac{1}{3}-\beta^2 + \frac{2}{3} \beta^3 \cos 3 \gamma}
\ .  \ee

We wish to stress that the last two terms in the expression of the potential
energy (\ref{3e2}) and in particular the $\gamma$-dependence arise from the
nontrivial gauge invariant factor in the integration measure (\ref{2e13}).
This potential diverges for $\beta=1, \gamma=0$.
Therefore, the wave function has to vanish at this point.
For $\gamma = 0$ and
$\beta=1-\epsilon$, we have, in the limit of small $\epsilon$ :
\be \label{3e2a}
\Psi \left( \rho, \beta, \gamma \right) = \sqrt 3 \:
\frac{\epsilon^{1/2}}{\rho^8} \: \Phi \left( \rho, \beta, \gamma \right)
\ . \ee

In the limit of small deformations $\beta \ll 1$, the vibrational kinetic
energy can be written as :
\be \label{3e3}
\ba{ll}
T_{vib} =  -\frac{1}{2 L^3} \:
& {\displaystyle
\left\{
\frac{\partial^2}{\partial \rho^2} \right. } \\
& {\displaystyle \left. +
\frac{2}{\rho^2} \left[ \frac{1}{\beta^4} \: \frac{\partial}{\partial \beta}
\left( \beta^4 \frac{\partial}{\partial \beta} \right) +
\frac{1}{\beta^2 \sin 3 \gamma} \: \frac{\partial}{\partial \gamma} \left(
\sin 3 \gamma \frac{\partial}{\partial \gamma} \right)
\right] \right\}  \ . } \ea
 \ee
We thus obtain in this limit the expression of Bohr and Mottelson for
the vibrational kinetic energy.

The next step is to find the equilibrium shape by minimizing the potential
energy $V' \left( \rho, \beta, \gamma \right)$ with respect to
$\rho, \beta, \gamma$.
For this purpose, we have calculated the derivatives of the potential
and the general expression for the stability matrix.

Spherical configurations ($\beta = 0$ and $ L^2 \bar \rho_0^2 = g^{-2/3}
\left( \frac{3^4}{2} \right)^{1/3}$ )
correspond to maxima with an energy equal to :
$\bar V' \left( \bar \rho_0, \beta_0, \gamma =0 \right) = 5.9 \:  g^{2/3}
/L $.

We have found a
minimum at $\gamma =0$, $\beta = \bar \beta$, $\rho = \bar \rho$, where
$\bar \beta$ and $\bar \rho$ are solutions of the following equations :
\be \label{3e4}
L^6 \: g^2 \: \bar \rho^6 = - \frac{9}{4 \left( \frac{1}{3} - \bar \beta^2
+ \frac{2}{3} \bar \beta^3 \right)} \: \left( 1 -
\frac{\left(1-\bar \beta^2 \right) \left( 1- \bar \beta \right) }
{\frac{1}{3} - \bar \beta^2 + \frac{2}{3} \bar \beta^3} \right)
\ , \ee
\be \label{3e5}
L^6 \: g^2 \: \bar \rho^6 = \frac{9}{8 \left( 1 - \bar \beta^2 \right)}
\: \left( 33 + \frac{1- \bar \beta^2}{\frac{1}{3} -\bar \beta^2 + \frac{2}{3}
\bar \beta^3} \right)
\ . \ee
The two functions given by the right hand sides of eq. (\ref{3e4}) and of
eq. (\ref{3e5}) intersect at $\bar \beta$ very near 1. By writing
$\beta = 1- \epsilon$ and by keeping the lowest order in $\epsilon$, we
find at the minimum
\be \label{3e5a}
\bar \epsilon = \frac{2}{33}
\ . \ee and
\be \label{3e6}
L^2 \bar \rho^2 = \left( \frac{3}{2} \right)^{2/3} \: \frac{1}{g^{2/3} \:
\bar \epsilon^{2/3}} = \left( \frac{99}{4} \right)^{2/3} \:
\frac{1}{g^{2/3}}
\ . \ee
Therefore the minimum of $V'$ corresponds to a strongly deformed field
configuration with axial symmetry. We obtain :
\be \label{3e6a}
 \bar V' \left( \bar \rho^2, \beta=1-\bar \epsilon , \gamma=0 \right) =
  g^{2/3} \: \frac{3}{2} \left(\frac{99}{4}^{2/3}\right)/L \ . \ee
The corresponding value of the magnetic field is :
\be \label{3e7}
B^2 = \frac{2}{3} \: g^2 \: \bar \epsilon \bar \rho ^4 \ee
or
\be \label{3e8}
L^4 B^2 = \left( \frac{99}{4} \right)^{1/3} \: g^{2/3}  = g L \bar \rho \ . \ee
This minimum associated to a nonvanishing value of the magnetic field
cannot  be described by a perturbative approach which constructs
a state around $B^2=0$.

In lowest order in $\bar \epsilon$ or in $1/\bar \epsilon$, the
eigenvalues of the stability matrix at the minimum are
\be \label{3e9}
\omega_{\rho}^2 = \frac{9}{2} \: \left( \left( \frac{3}{2} \right)^{2/3}
-\frac{1}{2} \left( \frac{3}{2} \right)^{-1/3} \right) \: g^{4/3} \:
\bar \epsilon^{1/3} \ , \ee
\be \label{3e10}
\omega_{\beta}^2 = \left(\frac{3}{2} \right)^{1/3} \: g^{2/3} \:
\frac{1}{\bar \epsilon^{7/3}} \ , \ee
\be \label{3e11}
\omega_{\gamma}^2=3 \left(\frac{3}{2} \right)^{1/3} \: g^{2/3} \:
\frac{1}{\bar \epsilon^{7/3}} \ , \ee
where we have used dimensionless quantities. To this order in
$\bar \epsilon$, the eigenvectors of the stability matrix are given by
the directions of $\delta \rho$, $\delta \epsilon$ and
$\delta \gamma \equiv \eta$. Therefore, the quadratic expansion of the
potential energy arround the minimum yields :
\be \label{3e12}
V'(\rho, \beta, \gamma) = \bar V'(\bar \rho, \bar \epsilon,0) +
\frac{1}{2} L^3 \omega_{\rho}^2 (\delta \rho)^2 +
\frac{1}{2} L \omega_{\beta}^2 (\delta \epsilon)^2 +
\frac{1}{2} L \omega_{\gamma}^2 \eta^2 \ . \ee

\section{Solutions of the rotation-vibration hamiltonian}

In analogy with the procedure used in nuclear physics,
it will be more convenient to
redefine the wave function according to :
\be \label{4e1}
\varphi \left( \rho, \beta, \gamma \right) = \vert \sin 3 \gamma \vert^{1/2} \:
\Phi \left( \rho, \beta, \gamma \right) = \rho^{11/2}
\frac{ \vert \sin 3 \gamma \vert^{1/2} }{ \left\vert \lambda_1 \lambda_2
\lambda_3
\right\vert^{1/2} } \: \Psi \left( \rho, \beta, \gamma \right) \ . \ee
After this rescaling, the differential operator $T_{vib}$ is transformed into
$\tilde T_{vib} + V_{add}$. Near the minimum,
$\rho = \bar \rho + \delta \rho$, $\beta = 1- \bar \epsilon -\delta \epsilon$,
$\gamma = \eta$, we have :
\be \label{4e2}
\tilde T_{vib} = - \frac{1}{2 L^3} \: \left\{ \frac{\partial^2}
{\partial \rho^2}
+ \frac{6 \bar \epsilon}{\bar \rho^2} \: \frac{\partial^2}{\partial \epsilon^2}
+ \frac{6}{\bar \rho^2} \: \frac{\partial}{\partial \epsilon} +
\frac{2 \bar \epsilon}{\bar \rho^2}
\: \frac{\partial^2}{\partial \eta^2} \right\}
\ , \ee
and
\be \label{4e3}
V_{add}= - \frac{1}{4 L^3} \: \frac{\bar \epsilon}{\bar \rho^2 \eta^2} \ . \ee
In $\tilde T_{vib}$, we have neglected terms proportional to
$\eta \: \frac{\partial}{\partial \eta}$.

The rotational kinetic energy in eq. (\ref{1e8}) can be written as :
\be \label{4e4}
T_{rot} = \frac{1}{2 L^3} \left[ \frac{L_1^2}{{\cal J}_1} +
\frac{L_2^2}{{\cal J}_2} + \frac{L_3^2}{{\cal J}_3} \right] \ . \ee
The rotational and the vibrational parts of the energy are coupled to each
other due to the $\rho$, $\beta$, $\gamma$ dependence of the moments of
inertia :
\be \label{4e5}
{\cal J}_1 = 2 \rho^2 \beta^2 \: \frac{\sin^2 \left( \gamma + 2 \pi/3 \right)}
{1- \beta \cos \left( \gamma + 2 \pi/3 \right)} \ , \ee
\be \label{4e6}
{\cal J}_2 = 2 \rho^2 \beta^2 \: \frac{\sin^2 \left( \gamma +  \pi/3 \right)}
{1+ \beta \cos \left( \gamma +  \pi/3 \right)} \ , \ee
\be \label{4e7}
{\cal J}_3 = 2 \rho^2 \beta^2 \: \frac{\sin^2  \gamma }
{1- \beta \cos  \gamma } \ . \ee

The collective hamiltonian in terms of the Euler angles and the variables
$\rho$, $\beta$, $\gamma$ is given by
\be \label{4e8}
H = T_{rot} + \tilde T_{vib}
+ V_{add} + V'\left( \rho, \beta, \gamma \right) \ . \ee
The wave functions are normalized according to the volume element
\hfill\break
$\beta^4  \: d \rho \: d \beta \: d \gamma
\: d \Omega$ where $d  \Omega = sin \theta_2 \: d \theta_1 \: d \theta_2 \:
d \theta_3$ is the volume element of the Euler angles.

In order to obtain approximate solutions for the dynamics, we expand the
potential energy arround the axial minimum according to eq. (\ref{3e12}).
Near the minimum, the moments of inertia are given by :
\be \label{4e9}
{\cal J}_1 = {\cal J}_2 \simeq \bar \rho^2  \ , \ee
and
\be \label{4e10}
{\cal J}_3 = \frac{2 \bar \rho^2}{\bar \epsilon} \: \eta^2 \ . \ee
This yields for the kinetic rotational energy :
\be \label{4e11}
T_{rot} = \frac{1}{L^3} \: \left( \frac{ L^2 - L^2_3}{2 \bar \rho^2 }
+ \frac{\bar \epsilon}{4 \bar \rho^2}
\: \frac{L_3^2}{\eta^2} \right) \ . \ee
The last term in $T_{rot}$ is a rotation-vibration interaction.
The presence of this centrifugal barrier has a deep connection with
the gauge invariance. It arises because when two $\lambda$'s are equal
we can not define the rotation angles in the polar representation
for $A_{ia}$. As a consequence, there is in this case
no  dynamics associated to
these degrees of freedom.

By keeping ${\cal J}_1$ and ${\cal J}_2$ constants, we will neglect further
rotation-vibration interaction terms. For small angular momentum, we can
treat them in perturbation. We have checked that they are indeed small,
 which is expected since the deformation  is large.

The eigenfunctions of H are of the form \cite{8a} :
\be \label{4e12} \ba{ll}
 \varphi \left(\delta \rho, \delta \epsilon, \eta , \theta_i \right) =
\left( \frac{2 I +1}{16 \pi^2 (1+ \delta_{K0})} \right)^{1/2} \:
& {\displaystyle
R_{n_{\rho}} \left(\delta  \rho \right) \: g_{n_{\beta}} \left( \delta
\epsilon \right) \: \chi_{n_{\gamma}}^K \left( \eta \right) \: } \\
& {\displaystyle \times
\left( D^{I*}_{MK} \left( \theta_i \right) + (-1)^I \: D^{I*}_{M-K}
\left( \theta_i \right) \right)  \ ,  } \ea \ee
where $R_{n_{ \rho}}$ are the solutions of the harmonic oscillator of frequence
$\omega_{\rho}$. For the functions $g_{n_{\beta}} \left( \delta \epsilon
\right)$ and $\chi^K_{n_{\gamma}} \left( \eta \right)$, we obtain the
following expressions :
\be \label{4e13}
g_{n_{\beta}} \left( \delta \epsilon \right)=
A \: \exp \left(- \frac{1}{4 \bar \epsilon^2} \left( \delta \epsilon + \bar
\epsilon \right)^2
\right)  \: H_{n_{\beta}}
\left( \frac{\delta \epsilon}{\sqrt 2 \bar \epsilon} \right)
\ , \ee
and
\be \label{4e13a}
\chi^K_{n_{\gamma}} (\eta)= B \: \vert \eta \vert^{1/2} \: \eta^{K/2} \:
\exp \left(- \frac{\lambda}{2} \eta^2 \right) \:
_1F_1 \left( -n_{\gamma}, \frac{\vert K \vert}{2} + \frac{1}{2},
\lambda \eta^2 \right) \ , \ee
where $H_{n_{\beta}}$ are the Hermite polynomials, $_1F_1$ is the
hypergeometric function and $\lambda=3/2\bar \epsilon^2$. A and B are
normalization constants.

We thus see from eq. (\ref{4e13})
that dynamical effects arising from the collective kinetic
energy shift the center of the gaussian away from the minimum of the
potential energy at $\beta =1-\bar \epsilon$ to the point $\beta=1$.
This point was already known to be the minimum of the magnetic energy
eq. (\ref{2e10}). However an additional information we have gained
from eq. (\ref{3e2}) is that the boundary condition on the wave function is
precisely that it must vanish at $\beta=1$.
This implies that only odd values of $n_{\beta}$ are acceptable.
A second boundary condition on the wave function is that it vanishes at
$\eta^2=\bar \epsilon^2/3$. For $\beta$ near 1 and $\gamma$ small,
we have from eq. (\ref{2e6}) and eq. (\ref{2e11}) :
$\left( \lambda_1 \: \lambda_2 \: \lambda_3 \right)^2 =
\bar \rho^6 \left( \bar \epsilon^2 - 3
\eta^2 \right)/9$. From the expression for $\chi^K_{n_{\gamma}} (\eta)$,
we see that the wave function is indeed negligible  at the boundary
$\eta^2 \simeq \bar \epsilon^2/3$.

The functions (\ref{4e12}) describe states of positive parity, K being
restricted to positive even integers and $I=0,1,2$ for $K=0$ and
$I=K, K+1$... for $K \ne 0$.

 The eigenvalues of the  energy are given by :
\be \label{4e14}
\ba{ll}
E_{IKn_{\rho}n_{\beta}n_{\gamma}} =
& {\displaystyle
\left(n_{\rho} + \frac{1}{2} \right)
\frac{1}{L} \omega_{\rho} + \left( n_{\beta} +\frac{3}{4} \right) \:
E_{\beta} + \left( 2 n_{\gamma} + \frac{1}{2} \vert K \vert + 1 \right) \:
E_{\gamma} } \\
& {\displaystyle +
\frac{1}{2 L^2 \bar \rho^2 } \:
\left( I(I+1) - K^2 \right) \frac{1}{L} + \bar V' \ , }  \ea
\ee
where
\be \label{4e15a}
E_{\beta}= E_{\gamma}  =  \frac{3}{L} \left(\frac{2}{3} \right)^{2/3}
\: g^{2/3} \:\frac{1}{ \bar \epsilon^{1/3}}  \ , \ee
and $\omega_{\rho}$ is given by (\ref{3e9}).

The $\rho$ vibration is therefore  softer than the $\beta$ and $\gamma$
vibrations.
Note that the zero point vibrational energy for the $\beta$ vibration is
$7/4 \: E_{\beta}$ instead of the usual value $1/2$.

In the strong coupling approximation, the energy is proportional to $g^{2/3}$.
By using $\bar \epsilon = 2/33$, one finds :
\be \label{4e16}
\omega_{\rho} = 1.24 \: g^{2/3}  \ , \ee
and
\be \label{4e18}
 E_{\beta} =  E_{\gamma} =  5.83 \: g^{2/3} /L \ . \ee

 The lowest excited energy levels are the states
 of the ground state
rotational band $K=0$ : $m(2^{+})= 0.35$ and $m(4^{+})= 1.18$,
the states of the $\rho$-vibrational band $n_{\rho}=1$ :
$m(0^{+}) = 1.24 , m(2^{+})=1.59$ and the state in the band $n_{\rho}=2$ :
$m(0^{+})=2.5$ (all the masses are given in units of $1/L$).

For the $\rho$ vibration, we have also investigated the importance of the
anharmonic terms.
At the lowest order in $\bar \epsilon$, the collective potential is given
by
\be \label{4e19}
V'(\rho)=g \: \rho + \frac{99}{8L^3 \rho^2} \ , \ee
or, using the appropriate rescaling $ \rho = g^{1/3} \: \tilde \rho/L$,
\be \label{4e20}
 V'(\tilde \rho)=
g^{2/3} \: \left(\tilde \rho + \frac{99}{8 \tilde \rho^2}
\right)\: \frac{1}{L} \ . \ee
The cubic and quartic terms give sizeable contributions of opposite signs.
If we approximate the collective potential by a quartic polynomial,
the mass of the first $0^+$ excited state is lowered to :
$ m(0^+) \simeq 1.0/L$.

To perform a comparison with experimental data it would be necessary to
generalize the present calculation to the SU(3) group and to include
dynamical quarks.
 The experimental situation however appears
somewhat unsettled \cite{10}.

\section{ Discussion and perspectives}

 In order to compare with the approaches which consider gauge invariant
variables constructed from the magnetic field
\cite{1,8b}, it is usefull to remember
 the following relations valid in the strong coupling
limit between the three
eigenvalues $b_n^2$ of the matrix $B_{ia}B_{ja}$ and the three $\lambda$'s :
\be \label{5e1}
b_1=g \lambda_2  \: \lambda_3 \: ,
b_2=g \lambda_1 \: \lambda_3 \: , b_3 = g \lambda_1 \: \lambda_2 \ . \ee
We have also $B . y = 6 \: b_1 b_2 b_3$.

Our result for the behaviour of the wavefunction in the infrared region
is in agreement with the conclusion of K. Johnson \cite{1}, who found that
the vacuum wave function should vanish at $\det B=0$.
The results $\bar \rho^2 \ne 0$ and $B^2 \ne 0$ at the minimum cannot
be obtained in a perturbative approach.
We have found a strongly deformed minimum and the lowest energy levels
correspond to a rotational band and a $\rho$-vibrational band.
Our predictions for the lowest
glueball states are   different from those of L\"uscher and M\"unster
who have performed a perturbative calculation for the $SU(2)$
 gauge theory in a finite volume \cite{6}.
They have choosen a basis of eigenfunctions proportional to
$\exp (-\frac{1}{2} \omega A_{ia}A_{ia})$ to
diagonalize their effective hamiltonian. This basis is not adapted
to describe our wave functions
(\ref{4e12}) : a very large number of terms would be necessary to have
a sufficient accuracy.  In contrast,
if we had obtained a minimum located at
$\bar \rho = 0$, a gaussian in the cartesian coordinates would have
been an acceptable ansatz \cite{6a}. Because the minimum is located at
$\bar \rho \ne 0$, the correct ansatz is a gaussian in the
"curvilinear coordinates" $\delta \rho$, $\delta \epsilon$ and $\eta$,
which cannot be expressed as a gaussian in the cartesian coordinates
$A_{ia}$.

Let us make  some remarks about the derivative expansion.
This is an expansion in powers of $1/g^{2/3}$. It is expected to
 be valid when the
derivatives of gauge invariant quantities are small compared to some
scale, for instance $\vert \partial_i \lambda_n \vert \ll \bar \rho^2 $.
In lowest order in the strong coupling approximation, there is no
propagation and the wavefunctional reduces to a function of
gauge invariant quantities constant in space.
 The next step is to investigate the coupling between the sites and
its effect on the vacuum state properties and the glueball spectrum.

For color singlet states of zero angular momentum, the first term involving
derivatives in the kinetic energy is :
\be \label{5e2}
T_1 = -\frac{1}{2g^2} \left\{ \sum_{n,m,k} \: \lambda_n \: S_{nm}^k \:
\left( \vec \nabla . \vec f^k \right) \frac{\delta}{\delta \lambda_k}
\: \lambda_n \: S_{nm}^k \: \left( \vec \nabla . \vec f^k \right)
\frac{\delta}{\delta \lambda_k} \right\} \ , \ee
where
\be
S_{nm}^k = -\frac{\epsilon_{knm}}{\lambda_n^2-\lambda_m^2} \ , \ee
and $\vec f^n$ are the column vectors of the matrix $\hat f$.
For the potential energy, the terms containing derivatives are
exactly given by the last term in eq. (\ref{1e7}).
In order to check the accuracy of our variational Ansatz,
it will be  usefull to
 compare our results with other
SU(2) calculations : finite volume results \cite{6},
 lattice Monte Carlo
calculations \cite{7} and analytical strong coupling expansions \cite{8}.
This comparison will also provide a usefull guide to construct variational
Ans\"atze for negative parity states which are not included  in our present
variational space. Indeed the rescaling we have performed implies trial
wave functions of the form $\Psi(\lambda_1^2, \lambda_2^2, \lambda_3^2)$.

\vspace{1cm}
{\bf Aknowledgements}
One of us (C. M.) thanks K. Johnson for very enlightening discussions.

\end{document}